\documentclass[preprint,review,12pt]{elsarticle}

\usepackage{extsizes}
\usepackage[version=3]{mhchem}
\usepackage[left=1.5cm, right=1.5cm, top=1.785cm, bottom=2.0cm]{geometry}
\usepackage{times,mathptmx}
\usepackage{sectsty}
\usepackage{epsfig}
\usepackage[english]{babel}
\usepackage[T1]{fontenc}
\usepackage[usenames,dvipsnames]{xcolor}
\usepackage[compact]{titlesec}
\usepackage{textcomp}
\usepackage{amssymb}
\usepackage{epstopdf}

\makeatletter
\def\ps@pprintTitle{%
     \let\@oddhead\@empty
     \let\@evenhead\@empty
     \def\@oddfoot{}%
		 \let\@evenfoot\oddfoot}
\makeatother

\begin{document}

\begin{frontmatter}

\title{Molecular mechanisms driving the microgels behaviour: a Raman spectroscopy and Dynamic Light Scattering study}

\author[1,2]{Valentina Nigro}
\author[2]{Francesca Ripanti \corref{corr1}}
\author[1,2]{Roberta Angelini \corref{corr1}}
\author[3]{Angelo Sarra}
\author[4]{Monica Bertoldo}
\author[5]{Elena Buratti}
\author[2]{Paolo Postorino \corref{corr1}}
\author[1,2]{Barbara Ruzicka}

\address[1]{Istituto dei Sistemi Complessi del Consiglio Nazionale delle Ricerche (ISC-CNR), sede Sapienza, Pz.le Aldo Moro 5, I-00185 Roma, Italy}
\address[2]{Dipartimento di Fisica, Sapienza Universit$\grave{a}$  di Roma, P.le Aldo Moro 5, 00185 Roma, Italy}
\address[3]{Dipartimento di Scienze, Universit$\grave{a}$ degli Studi Roma Tre, via della Vasca Navale 84, 00146 Roma, Italy}
\address[4]{Istituto per la Sintesi Organica e la Fotoreattivit$\grave{a}$ del Consiglio Nazionale delle Ricerche (ISOF-CNR), via P. Gobetti 101, 40129 Bologna, Italy}
\address[5]{Istituto per i Processi Chimico-Fisici del Consiglio Nazionale delle Ricerche (IPCF-CNR), Area della Ricerca, Via G.Moruzzi 1, I-56124 Pisa, Italy}

\cortext[corr1]{Corresponding authors: francesca.ripanti@uniroma1.it (F.Ripanti); roberta.angelini@roma1.infn.it (R.Angelini); paolo.postorino@roma1.infn.it (P.Postorino)}

\begin{abstract}
Responsive microgels based on poly(N-isopropylacrylamide) (PNIPAM) 
exhibit peculiar behaviours due to the competition between the hydrophilic and hydrophobic 
interactions of the constituent networks. The interpenetration of poly-acrilic acid (PAAc), a pH-sensitive polymer,
within the PNIPAM network, to form Interpenetrated Polymer Network (IPN) 
microgels, affects this delicate balance and the typical Volume-Phase Transition (VPT) 
leading to complex behaviours whose molecular nature is still completely 
unexplored.

Here we investigate the molecular mechanism driving the VPT and its 
influence on particle aggregation for PNIPAM/PAAc IPN microgels by the joint use
of Dynamic Light Scattering and Raman Spectroscopy. Our results 
highlight that PNIPAM hydrophobicity is enhanced by the interpenetration of PAAc promoting 
interparticle interactions, a  crossover concentration is 
found above which aggregation phenomena become relevant.
Moreover we find that, at variance with PNIPAM, for IPN microgels a double-step molecular mechanisms occurs upon crossing the VPT, the first involving the coil-to-globule transition typical of PNIPAM and the latter associated to PAAc steric hindrance. 
\end{abstract}


\end{frontmatter}

\section{Introduction}
\label{Introduction}

Responsive microgels are highly attractive systems for several technological applications due to their high responsiveness to slight changes of environmental conditions \cite{LyonRevPC2012}. Smart microgels have indeed many applications in agriculture, construction, cosmetics and pharmaceutics 
industries, in artificial organs, and tissue engineering \cite{ VinogradovCurrPharmDes2006, DasAnnRevMR2006, ParkBiomat2013, HamidiDrugDeliv2008, SmeetsPolymSci2013, SuBiomacro2008}. Moreover they are widely investigated in fundamental physics as good model systems for understanding the intriguing behaviours of soft colloids \cite{LikosJPCM2002,RamirezJPCM2009,HeyesSM2009}. Their interparticle potential and their effective volume fraction can be indeed modulated through easily accessible control parameters driving the system to unconventional phase-behaviours \cite{WangChemPhys2014, PritiJCP2014, HellwegCPS2000, WuPRL2003}, drastically different from those of conventional hard colloidal systems \cite{PuseyNat1986, ImhofPRL1995, PhamScience2002, EckertPRL2002, LuNat2008, RoyallNatMat2008, RuzickaNatMat2011, AngeliniNC2014}. 

In the last years, poly(N-isopropylacrylamide) (PNIPAM) has become very popular due to its thermo-sensitivity \cite{PeltonColloids1986}: at room temperature the polymer is hydrophilic and strongly hydrated in solution, while it becomes hydrophobic above 305 K, leading to a coil-to-globule transition that gives rise to a Volume-Phase Transition (VPT) from a swollen to a shrunken state~\cite{WuMacromol2003} largely studied both experimentally \cite{NigroCSA2017, StiegerMacromol2009, KratzPolymer2001} and theoretically  \cite{Flory1953, LeleMacromol1998, HinoJAPS1996, OtakeMacromol1990, LopezLeonPRE2007}. 
The typical swelling/shrinking transition affects particle interactions and drives the system towards unusual phase behaviours \cite{MattssonNature2009, PaloliSM2013, LyonRevPC2012, PritiJCP2014, WangChemPhys2014}, controlled by changing concentration \cite{TanPolymers2010,WangChemPhys2014},  solvents \cite{ZhuMacroChemPhys1999}, internal structure and composition (such as number and distribution of cross-linking points \cite{KratzBerBunsenges11998, KratzPolymer2001} and core-shell structure~\cite{HellwegLangmuir2004, MengPhysChem2007}) or by introducing additives into the PNIPAM network \cite{HellwegLangmuir2004}.

This mechanism can be even more complex if pH-sensitive polymers are introduced within the PNIPAM network, allowing to tune the polymer/polymer and polymer/solvent interactions. In particular the introduction of poly-acrilic acid (PAAc) to the PNIPAM microgel permits to control the temperature dependence of the VPT by pH \cite{KratzColloids2000, KratzBerBunsenges21998, XiaLangmuir2004, JonesMacromol2000, NigroJNCS2015, NigroJCP2015}, PAAc content \cite{HuAdvMater2004, MaColloidInt2010} or ionic strength \cite{KratzColloids2000, XiongColloidSurf2011}.
The response of PNIPAM-PAAc microgels is strictly related  to the mutual interference between the two monomers (Fig. \ref{fig:mol}) \cite{KratzColloids2000, KratzBerBunsenges21998, JonesMacromol2000, XiongColloidSurf2011,
MengPhysChem2007, LyonJPCB2004, HolmqvistPRL2012, DebordJPCB2003}. If the PAAc is interpenetrated into PNIPAM to obtain PNIPAM/PAAc Interpenetrated Polymer Network (IPN) microgel,~\cite{HuAdvMater2004, XiaLangmuir2004, XiaJCRel2005, ZhouBio2008, XingCollPolym2010, LiuPolymers2012} the coil-to-globule transition temperature of PNIPAM  is almost unchanged \cite{MaColloidInt2010}, since irreversible chemical bonds between PNIPAM and PAAc chains are mainly excluded. This leaves the VPT temperature almost the same for PNIPAM and IPN microgels. 
Up to now a microscopic interpretation of the swelling behaviour for linear PNIPAM derives from the Flory Rehner theory that recently applied to PNIPAM based microgel under proper approximations \cite{QuesadaPerezSM2011, NigroCSA2017}. However, despite the numerous investigations on the VPT of PNIPAM microgels, a clear picture of the molecular mechanism behind swelling is still missing. 
To this aim Raman spectroscopy represents a powerful tool to highlight molecular changes related to the swelling behaviour as previously reported for  linear PNIPAM \cite{Dybal2009, FutscherSciRep2017} and PNIPAM microgels \cite{AhmedJPCB2009, WuJPCB2018}. The main outcomes of these investigations underline that the principal groups involved in the swelling are the CH$_2$ stretching bands of the methylene group and the CH$_3$ stretching bands of the isopropyl group (Fig. \ref{fig:mol}).
In the case of IPN microgels the presence of PAAc leads to more complex behaviour whose molecular nature has never been investigated up to now. 

In the present work we report on a careful study for PNIPAM and IPN microgels combining Dynamic Light Scattering (DLS) and Raman spectroscopy.
The interpretation of the experimental results allowed us to recognize different microscopic behaviours for the two systems. In particular, at variance with PNIPAM, we identify two molecular mechanisms driving the volume phase transition: one
responsible for the coil-to-globule transition of the PNIPAM and the other ascribed to the PAAc steric hindrance.

\section{Experimental Methods}
\label{Experimental Methods}

\subsection{Sample preparation}

\paragraph*{Materials}
N-isopropylacrylamide (NIPAM) (Sigma-Aldrich), purity 97 \%, and N,N'-methylene-bis-acrylamide (BIS) (Eastman Kodak), electrophoresis grade, were purified by recrystallization from hexane and methanol, respectively, dried under reduced pressure (0.01 mmHg) at room temperature and stored at 253 K. Acrylic acid (AAc) (Sigma-Aldrich), ), purity 99 \%, with 180 - 220 ppm of MEHQ as inhibitor, was purified by distillation (40 mmHg, 337 K) under nitrogen atmosphere on hydroquinone and stored at 253 K. Sodium dodecyl sulphate (SDS), purity 98 \%, potassium persulfate (KPS), purity 98 \%, ammonium persulfate (APS), purity 98 \%,  N,N,N',N'-tetramethylethylenediamine (TEMED), purity 99 \%, ethylenediaminetetraacetic acid (EDTA), purity $\geq$ 98.5 \%, NaHCO$_{3}$, purity 99.7 - 100.3 \% were all purchased from Sigma-Aldrich and used as received. Ultrapure water (resistivity: 18.2 M$\Omega$ $\cdot$ cm at 298 K) was obtained with Millipore Direct-Q\textsuperscript{\textregistered} 3 UV purification system.  All other solvents were RP grade (Sigma Aldrich) and were used as received. Before use, dialysis tubing cellulose membrane, cut-off 14,000 Da, from Sigma-Aldrich, was washed for 3 h in running distilled water, treated at 343 K for 10 min into a solution containing a 3.0 \% weight concentration of NaHCO$_{3}$ and 0.4 \% of EDTA, rinsed in distilled water at 343 K for 10 min and finally in fresh distilled water at room temperature for 2 h.

\paragraph*{Synthesis of PNIPAM and IPN microgels}
PNIPAM micro-particles were synthesized by precipitation polymerization with (24.162 $\pm$ 0.001) g of NIPAM, (0.4480 $\pm$ 0.0001) g of BIS and (3.5190 $\pm$ 0.0001) g of SDS solubilized in 1560 mL of ultrapure water and transferred into a 2000 mL four-necked jacked reactor equipped with condenser and mechanical stirrer. The solution was deoxygenated by bubbling nitrogen inside for 1 h and then heated at (343.0 $\pm$ 0.1) K. (1.0376 $\pm$ 0.0001) g of KPS (dissolved in 20 mL of deoxygenated water) was added to initiate the polymerization and the reaction was allowed to proceed for 16 h. The resultant PNIPAM microgel was purified by dialysis against distilled water with frequent water change for 2 weeks.
In the second step IPN microgels were synthesized by a sequential free radical polymerization method \cite{XiaLangmuir2004} with (140.08 $\pm$ 0.01) g of the PNIPAM dispersion at the final weight concentration of 1.06 \%. 5 mL of AAc was copolymerized with (1.1081 $\pm$ 0.0001) g of BIS into the preformed PNIPAM microparticles at temperature in the range 293$\div$295 K, where PNIPAM particles are swollen allowing the growth of the PAAc network inside them. The mixture was diluted with ultrapure water up to a volume of 1260 mL and transferred into a 2000 mL four-necked jacketed reactor kept at (294 $\pm$ 1) K by circulating water and deoxygenated by bubbling nitrogen inside for 1 h. 0.56 mL of TEMED were added and the polymerization was started with (0.4441 $\pm$ 0.0001) g of ammonium persulfate and allowed to proceed for 4 h and 30 min. The obtained IPN microgel at the final PAAc concentration of  $C_{PAAc}$ = 23 \% was purified by dialysis against distilled water with frequent water change for 2 weeks and then lyophilized up to 1.00 \% weight concentration. Samples at different weight concentrations, in the following referred as $C_w$, were obtained by dilution in H$_2$O.

 \subsection{Dynamic Light Scattering}
 \label{Dynamic Light Scattering}

Dynamic Light Scattering measurements have been performed with a light
scattering setup, where a monochromatic and polarized beam emitted from a solid state laser (100 mW at $\lambda$=642 nm) is focused on the sample placed in a cylindrical VAT for index matching and temperature control. The scattered
intensity is collected by single mode optical fibers at fixed scattering
angles, namely $\theta$=90\textdegree, corresponding to $Q=0.018$ $nm^{-1}$, according to the relation
Q=(4$\pi$n/$\lambda$) sin($\theta$/2).
The information on the system dynamics are extrapolated from the normalized intensity autocorrelation function
$g_2(Q,t)=<I(Q,t)I(Q,0)>/<I(Q,0)>^{2}$, directly measured through DLS with a high coherence factor close to the ideal unit value.
Measurements have been performed on aqueous suspensions of PNIPAM and IPN microgels at fixed PAAc content ($C_{PAAc}=23 \%$) as a function of temperature in the range T=(293$\div$313) K, at four weight concentrations ($C_w$=0.1 \%, $C_w$=0.3 \%, $C_w$=0.5 \% and $C_w$=0.8 \%) and pH 5.5. Reproducibility has been tested by repeating measurements several times.

As usual for colloidal systems, the intensity correlation functions are well described by the Kohlrausch-Williams-Watts
expression~\cite{KohlrauschAnnPhys1854, WilliamsFaradayTrans1970}:

\begin{equation}
g_2(Q,t)=1+b[(e^{-t/\tau})^{\beta}]^{2} \label{Eqfit}
\end{equation}

where $b$ is the coherence factor, $\tau$ is an "effective" relaxation time, defining the decay constant $\Gamma(Q)=1 / \tau(Q)$, and $\beta$ describes the deviation from the simple exponential decay ($\beta$ = 1) usually found in monodisperse systems and gives a measurement of the relaxation times distribution due to the intrinsic sample polydispersity. Many glassy materials show a stretching of the correlation functions (here referred to as "stretched behaviour") characterized by an exponent $\beta$ < 1.
Hydrodynamic radii R$_H$ have been determined  from the decay constant
$\Gamma(Q)=D q^2$ obtained through the analysis of the intensity correlation functions g$_2(Q,t)$ in the high dilution limit ($C_w$=0.1 \%). 

\subsection{Raman Spectroscopy}
\label{Raman}

Raman measurements have been carried out using a Horiba HR-Evolution microspectrometer in backscattering geometry, equipped with a He-Ne laser, $\lambda=632.8 \: nm$ and $30 \: mW$ output power ($\sim 15 \: mW$ at the sample surface). The elastically scattered light was removed by a state-of-the-art optical filtering device based on three BragGrate notch filters  \cite{Glebov2012} which also allows to collect Raman spectra at very low frequencies (down to $10 \:{cm}^{-1}$ from the laser line). The detector was a Peltier-cooled charge-coupled device (CCD) and the resolution was better than $3 \: {cm}^{-1}$ thanks to a $600 \: grooves/mm$ grating with $800 \:mm$ focal length.  The spectrometer was coupled with a confocal microscope supplied with a set of interchangeable objectives with long working distances and different magnifications ($20x$ - $0.35 \: NA$ was used for the present experiment). 
Further details on the experimental apparatus can be found in \cite{Capitani2015}.
Measurements have been performed on aqueous suspensions of PNIPAM and IPN microgels at fixed PAAc concentration ($C_{PAAc}$=23 \%) in the temperature range T=(293$\div$313) K across the VPT, at two weight concentrations ($C_w$=0.3 \% and $C_w$=5.0 \%) and pH 5.5.
Note that several measurements have been performed at lower weight concentrations resulting in a low signal-to-noise ratio that gives origin to misleading determination of the peak frequency.

\section{Results and Discussions}
\label{Results}

\begin{figure}[t]
\centering
\includegraphics[trim={0 0 0 0},clip, height=12cm, angle=90]{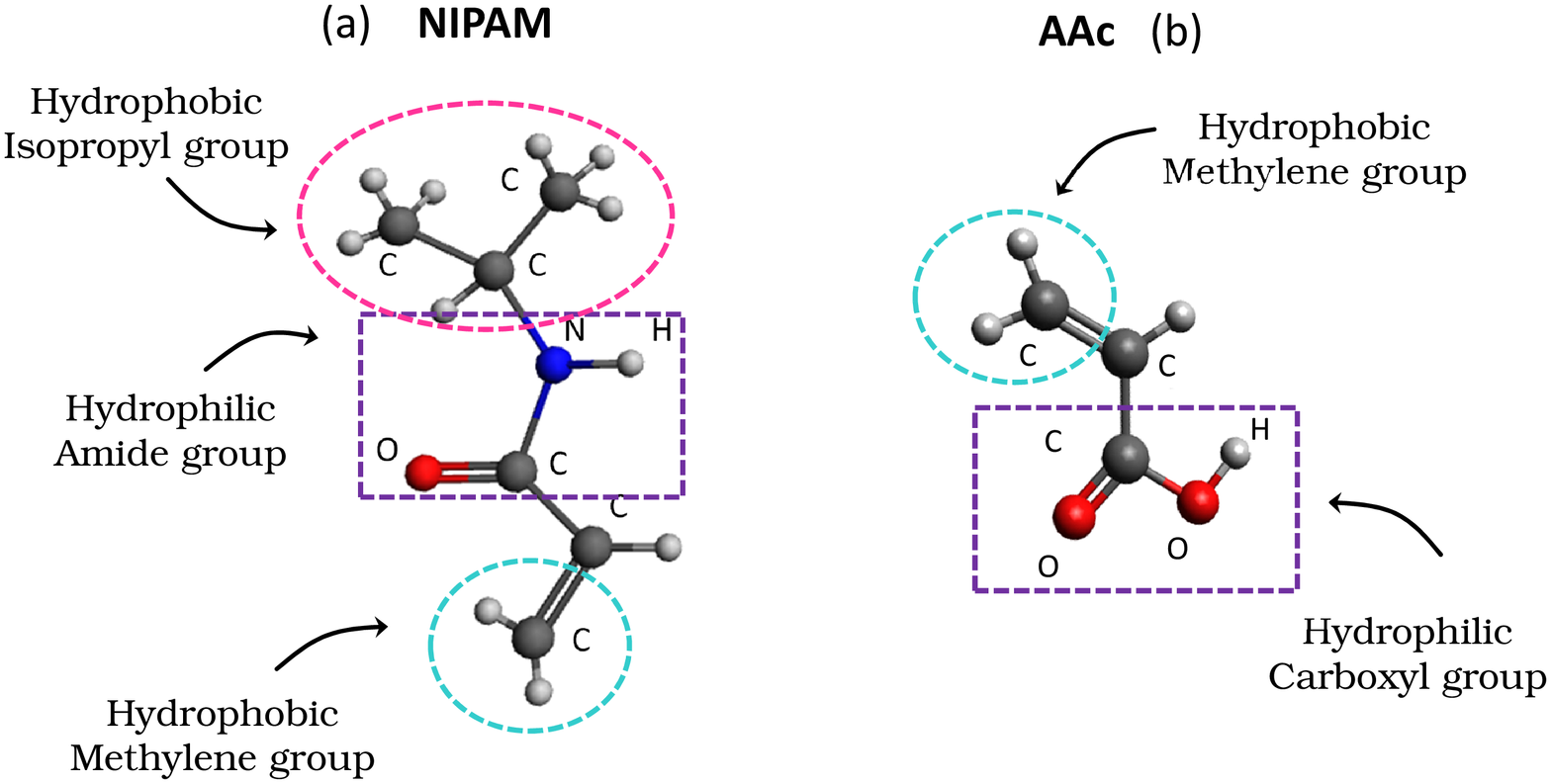}
\caption{Sketch of (a) NIPAM and (b) AAc molecular structure.}
\label{fig:mol}
\end{figure}

In order to connect the dynamical response to molecular changes of PNIPAM and IPN microgels across the VPT, we combine Dynamic Light Scattering and Raman spectroscopy measurements. 

\begin{figure}[t]
\centering
\includegraphics[height=10cm]{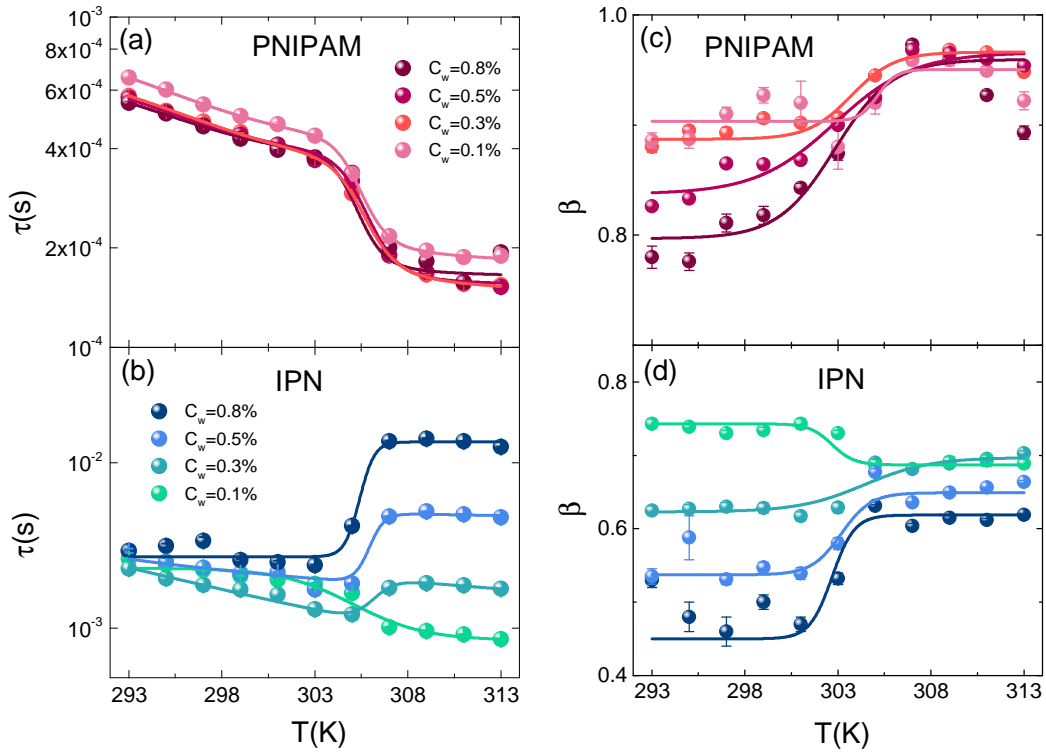}
\caption{(a,b) Relaxation time and (c,d) shape parameter as a function of temperature for PNIPAM and IPN microgels at $C_{PAAc}$=23 \%  at the indicated weight concentrations. Solid lines are guides to eyes.}
\label{fig:tau}
\end{figure}

The relaxation time and the stretching parameter for both PNIPAM and IPN microgels are reported in Fig.\ref{fig:tau} as a function of temperature at four weight concentrations (C$_w$=0.1 \%, C$_w$=0.3 \%, C$_w$=0.5 \% and C$_w$=0.8 \%) as obtained by fitting the $g_2(Q,t)$ with Eq.(\ref{Eqfit}). 
In the case of pure PNIPAM (Fig.\ref{fig:tau}(a)) the well known dynamical transition associated to the VPT is evidenced  \cite{NigroJNCS2015, NigroCSA2017}: as temperature increases, the relaxation time $\tau$ slightly decreases up to the volume phase transition temperature (VPTT), above which it decreases to its lowest value, corresponding to the shrunken state and indicating a fastening of the dynamics related to the reduced size of particles. 

In the case of IPN microgels a different scenario shows up (Fig.\ref{fig:tau}(b)) and the temperature dependence of the relaxation time above the VPTT is strongly affected by weight concentration: while at low $C_w$ it is similar to that of pure PNIPAM, for C$_w\geq$0.3 \% it is reversed. The sudden increase above the VPTT indicate a slowing down of the dynamics and the formation of aggregates. 
The presence of PAAc, in fact, affects the delicate balance between hydrophobic and hydrophilic interactions. In particular, at this pH conditions (pH 5.5), where the fraction of deprotonated AAc moieties (COO$^-$, Fig.\ref{fig:mol}(b)) is small but not negligible, the collapse of the PNIPAM network above the VPTT is supposed to favor the exposure of PAAc dangling chains and in turn interparticle interactions.
The dynamical transition associated to the VPT is also observed in the temperature behaviour of the stretching parameter $\beta$ for PNIPAM and IPN microgels reported in Fig.\ref{fig:tau}(c),(d).
Moreover in the case of IPN intriguing differences depending on concentration are observed: for C$_w$<0.3 \% $\beta$ decreases upon crossing the VPTT, while for C$_w\geq$0.3 \% microgel collapse leads to an increase of the stretching coefficient that
gets more pronounced with increasing concentration. Both $\tau$ and $\beta$ behaviours sign the existence of a crossover concentration C$_w$=0.3 \% in IPN microgels at $C_{PAAc}$=23 \% above which interparticle interactions become important giving rise to aggregation.

\begin{figure}[t]
\centering
\includegraphics[height=9cm]{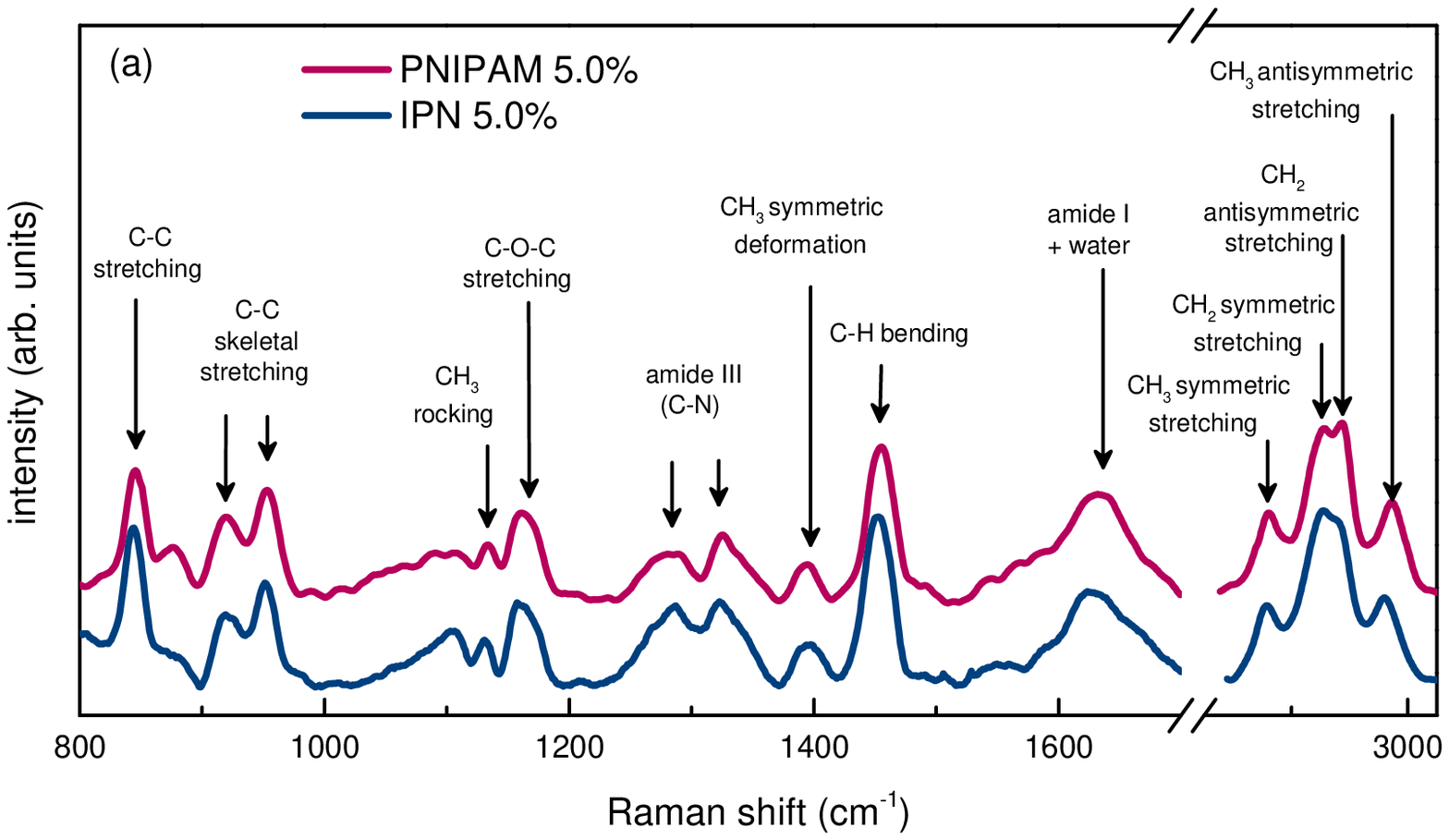}
\caption{Raman spectra for PNIPAM and IPN microgels at fixed temperature (T=316 K) and concentration (C$_w$=5.0  \%). }
\label{fig:spectraLab}
\end{figure} 

To deeply investigate the molecular mechanism driving the VPT and the conformational changes induced by microgel collapse, Dynamic Light Scattering measurements have been complemented with Raman spectroscopy on PNIPAM and IPN microgels at weight concentration equal and above the critical value C$_w$=0.3 \%.
Raman spectra over the frequency range from 800 to 3000 cm$^{-1}$ for both PNIPAM and IPN microgels are reported in Fig.\ref{fig:spectraLab} together with the assignment of the principal vibrational modes.
Interestingly the spectra are dominated by the contribution associated to the C-C and C-H vibration bands, mainly derived from NIPAM (Fig.\ref{fig:mol}(a)).

\begin{figure}[t]
\centering
\includegraphics[height=7cm]{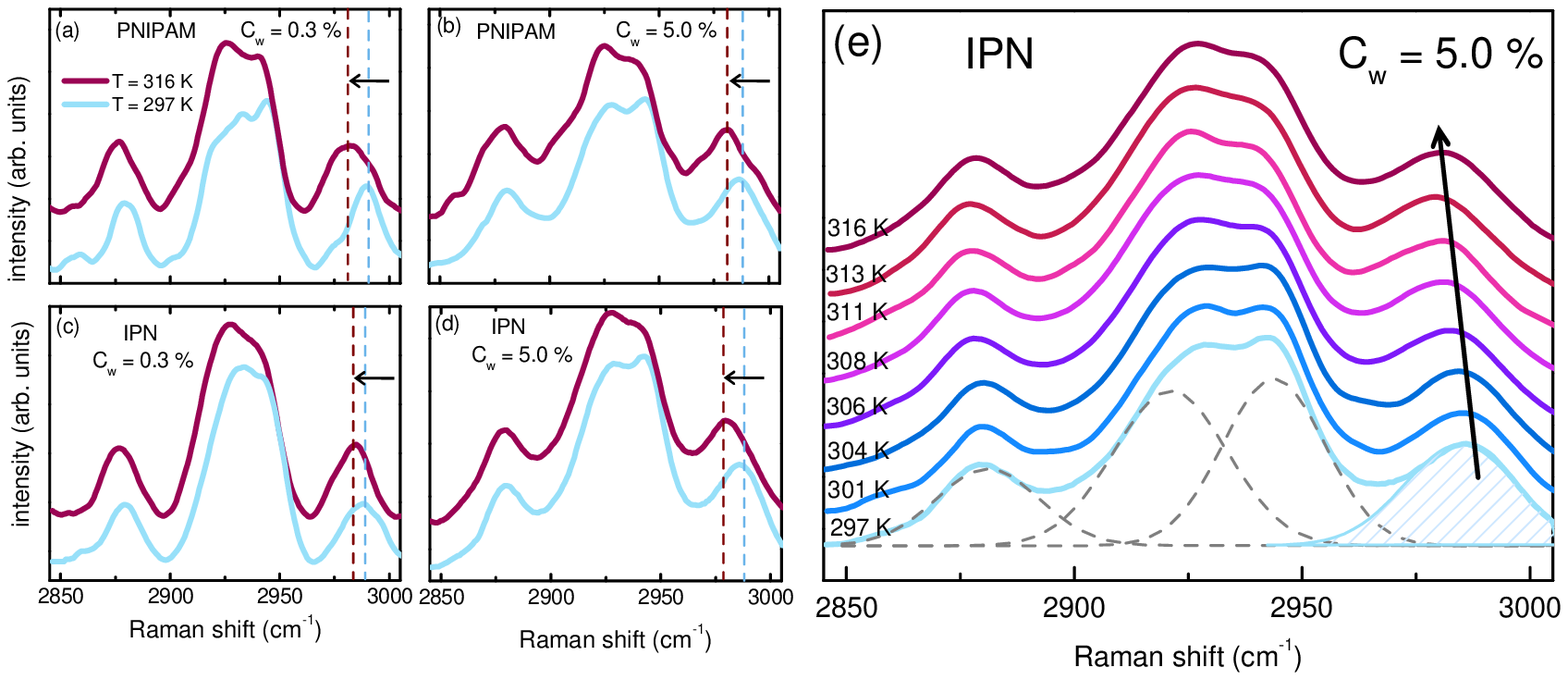}
\caption{Raman spectra for (a)(b) PNIPAM and (c)(d) IPN microgels at temperature below (light cyan) and above (red) the VPTT, at $C_w$=0.3 \% and $C_w$=5.0  \%. (e) Raman spectra for IPN microgels at $C_w$=5.0  \% at the indicated temperatures. Four Gaussian contributions are reported as discussed in the text. Dashed lines and arrows highlight the frequency of CH$_3$ stretching band.}
\label{fig:spectra}
\end{figure} 

To investigate the molecular changes induced by temperature across the VPT, we focus our attention on the spectral range between 2850 cm$^{-1}$ and 3000 cm$^{-1}$, where bands ascribed to vibrations of CH$_2$ (methylene group, Fig.\ref{fig:mol}(a)(b)) and CH$_3$ (in isopropyl group, Fig.\ref{fig:mol}(a)) are present. As recently observed in PNIPAM \cite{WuJPCB2018}, changes of these bands, sensitive to hydrogen bond variations, are related to different interactions both among polymers and between polymer side groups and water molecules surrounding them. 
Raman spectra of PNIPAM and IPN microgels at two concentrations $C_w$=0.3 \% (critical concentration) and $C_w$=5.0 \% are reported in Fig.\ref{fig:spectra} at temperatures T=297 K and T=316 K, below and above the VPTT (Fig.\ref{fig:spectra}(a,b,c,d)). The entire temperature behaviour upon crossing the VPT is reported in Fig.\ref{fig:spectra}(e) for IPN microgels at $C_w$=5.0  \% as an example. Spectra have been deconvolved by four Gaussian contributions (Fig.\ref{fig:spectra}(e)) that have been assigned to different C-H stretching of the NIPAM molecule in the hydrated state (T=297 K), according to previous works in literature \cite{Dybal2009, Tsuboi2008}: symmetric stretching of CH$_3$ (2880 cm$^{-1}$), symmetric and antisymmetric stretching of CH$_2$ (2920 cm$^{-1}$, 2945 cm$^{-1}$ respectively), antisymmetric stretching of CH$_3$ (2988 cm$^{-1}$).

Looking at Fig.\ref{fig:spectra}(a) and Fig.\ref{fig:spectra}(b), we notice a significant transfer of the spectral weight between the two central peaks ascribed to symmetric and antisymmetric CH$_2$ stretching modes of the methylene group. It is well known that the intensity ratio between the symmetric and antisymmetric stretching modes (lower and higher frequency, respectively) is related to the lateral packing density of any polymer chain \cite{TsuboiPolymJ2008}, in particular an increase of their ratio implies the folding of the linear chain. Therefore in our case the increase of the intensity ratio upon crossing the VPTT in both PNIPAM and IPN microgels (Fig.\ref{fig:spectra}(a,b,c,d)) can be ascribed to an increase of the packing density due to the coil-to-globule transition of PNIPAM signaling that their shrinking still drives the molecular changes across the VPT even when the PAAc is interpenetrated within the PNIPAM network.

\begin{figure}[t]
\centering
\includegraphics[height=7cm]{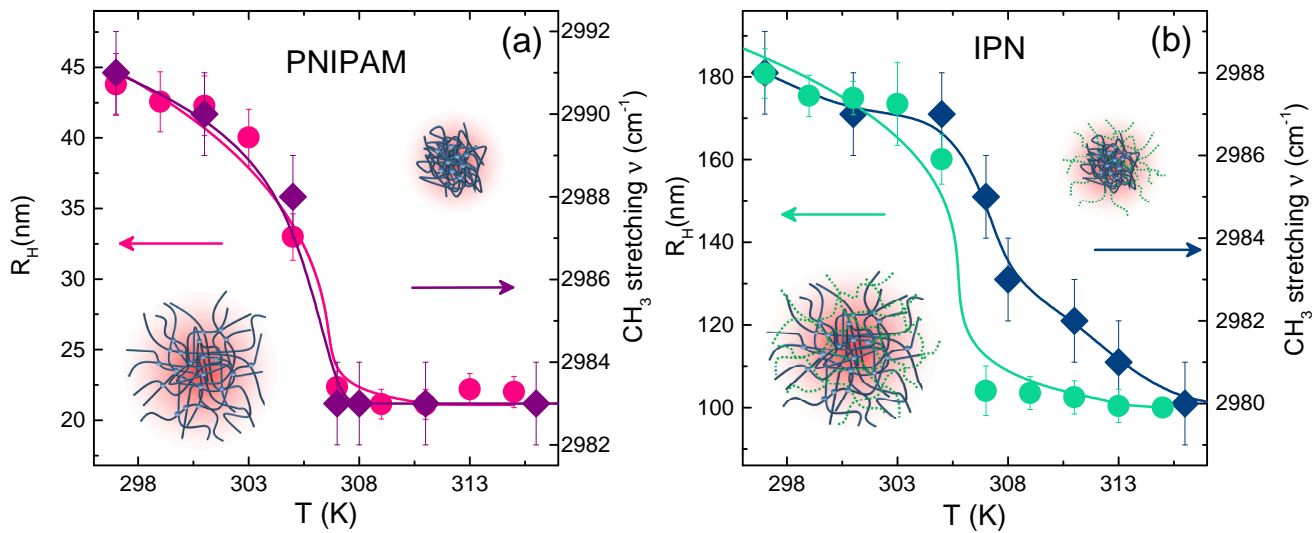}
\caption{Hydrodynamic radius R$_H$ and CH$_3$ frequency as a function of temperature for (a) PNIPAM and (b) IPN microgels at low concentrations. Magenta and green solid lines are the fit of R$_H$(T) through the Flory-Rehener theory, purple and blue solid lines are guide to eyes for the CH$_3$ frequency behaviour.}
\label{fig:CH3_03}
\end{figure}

Additional information on the swelling behaviour can be gained from the temperature dependence of the antisymmetric CH$_3$ stretching (higher frequency peak at 2988 cm$^{-1}$) \cite{Dybal2009}. 
A clear frequency red-shift upon crossing the VPTT is observed. This downshift with temperature (Fig.\ref{fig:spectra}) is the main signature of the dehydration of the isopropyl group of NIPAM (Fig.\ref{fig:mol}(a)). In fact, as previously observed and theoretically investigated, the higher number of water molecules surrounding the CH$_3$ groups correlates with the higher frequency of the CH$_3$ stretching vibration \cite{Schmidt2006, Hobza2002}. 
The frequency peak for PNIPAM microgels at $C_w$=0.3 \% is reported in Fig.\ref{fig:CH3_03}(a) as a function of temperature and compared with the hydrodynamic radius $R_H$ obtained from DLS. Both quantities show actually an identical temperature behaviour: they monotonically decrease as temperature increases showing a sharp transition close to the VPTT. This behaviour endorses the idea that the volume phase transition is accompanied by a reorganization of the neighboring water molecules leading to significant decrease of the hydration of the methyl group.
 
For IPN microgels the scenario is more complex since the presence of PAAc affects the balance between hydrophilic and hydrophobic interactions and the net charge. The decrease with temperature of the CH$_3$ frequency and of the hydrodynamic radius $R_H$, reported in Fig.\ref{fig:CH3_03}(b) for IPN microgels at $C_w$=0.3 \%, confirms that the main features of the coil-to-globule transition of PNIPAM are preserved. Moreover the smoother decrease with respect to PNIPAM is accompanied by an additional bump upon crossing the VPTT, suggesting the presence of additional molecular mechanisms due to the interpenetration of PAAc network. This two-steps decay can be explained in terms of a reduced hydration of the isopropyl groups of NIPAM (first drop) accompanied by a new mechanism (bump) due to the presence of PAAc that with its steric hindrance limits the microgel shrinking.

\begin{figure}[t]
\centering
\includegraphics[height=7cm]{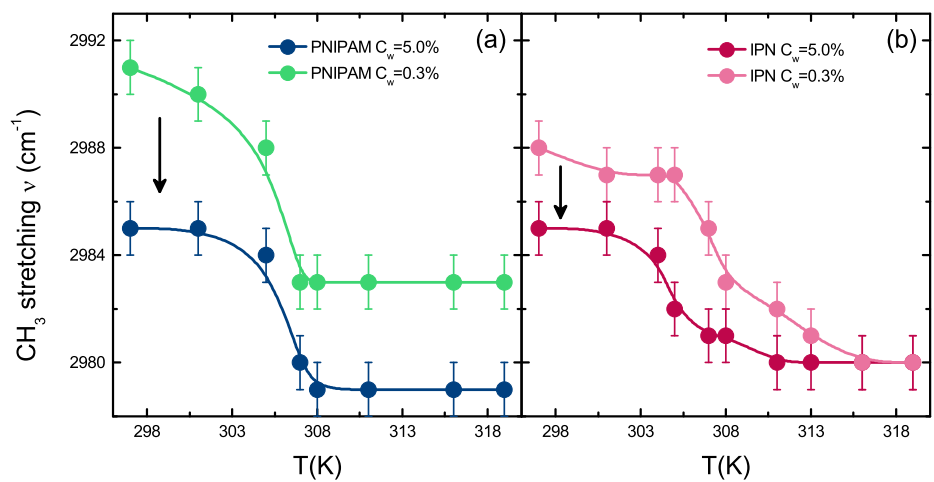}
\caption{CH$_3$ frequency as a function of temperature for (a) PNIPAM and (b) IPN microgels at $C_w$=0.3 \% and $C_w$=5.0  \%. Solid lines are guides to eyes. The arrows underline the reduction of CH$_3$ frequency on increasing microgel concentration.}
\label{fig:CH3_both}
\end{figure}

The behaviours of the CH$_3$ stretching frequency $\nu (T)$ for PNIPAM and IPN microgels at two concentrations ($C_w$=0.3  \% and $C_w$=5.0  \%) are reported in Fig.\ref{fig:CH3_both}(a) and Fig.\ref{fig:CH3_both}(b), respectively. The decrease of the frequency at higher concentration indicates a reduction of water molecules surrounding the isopropyl group corresponding to an higher internal packing density.
For IPN microgels (Fig.\ref{fig:CH3_both}(b)) the bump above mentioned becomes more evident at higher concentration ($C_w$=5.0  \%), suggesting that the C-H bonds of the PNIPAM methyl groups get stronger as particles become closer.  

A comparison between Fig.\ref{fig:CH3_both}(a) and Fig.\ref{fig:CH3_both}(b) highlights that for $C_w$=0.3  \% lower values of CH$_3$ frequency are observed in IPN with respect to PNIPAM microgels with a consequent smaller number of water molecules surrounding the PNIPAM side chains (CH$_3$ groups) in IPN and an higher internal packing density that we attribute to the interpenetration of the PAAc network. 
These results suggest that most of the differences between PNIPAM and IPN microgels are strictly related to the combined effect of reduced hydration of the CH$_3$ groups of PNIPAM and to the topological rearrangements of the polymer networks within the microgel particle.
Interpenetrating PAAc within PNIPAM enhances the hydrophobicity of the microgel particles perturbing the role played by water molecules. At these pH conditions (pH 5.5) the intra-particle and inter-particle interactions between CONH (PNIPAM) and COOH (PAAc) groups make respectively IPN microgel more hydrophobic and favor aggregation.
The same comparison for $C_w$=5.0  \%  (Fig.\ref{fig:CH3_both}(a) and Fig.\ref{fig:CH3_both}(b)) evidences that the initial and final values of the frequency are not significantly different for PNIPAM and IPN microgels indicating that for this concentration a dehydration limit has been reached independently of the system.

In summary, for PNIPAM microgel the VPT is characterized by a decrease of the relaxation time (Fig.\ref{fig:tau}(a)) and of the CH$_3$ frequency (Fig.\ref{fig:CH3_03}(a)) that reflect the rearrangement of water molecules as expected for swelling phenomenon. In the case of IPN microgel for C$_w\geq$ 0.3 \% the increase of the relaxation time above the VPTT (Fig.\ref{fig:tau}(b)), ascribed to aggregation phenomena, is associated to a more hydrophobic nature with respect to pure PNIPAM microgels, as also reflected in the temperature dependence of CH$_3$ frequency (Fig.\ref{fig:CH3_03}(b) and Fig.\ref{fig:CH3_both}(b)). Therefore we can argue that this novel feature can be considered a distinctive signature of IPN microgels.

\section{Conclusions}
\label{Conclusions}

The temperature behaviour across the VPT in PNIPAM and IPN microgels has been 
investigated through DLS and Raman spectroscopy. DLS has provided direct information about the hydrodynamic radius behaviour from a swollen to a shrunken state, whereas Raman spectroscopy has allowed to 
detect molecular modifications upon crossing the VPT not detectable through DLS. Surprisingly, for PNIPAM microgels, DLS and Raman spectroscopy yield 
equivalent temperature behaviours of completely different physical 
quantities. 
At variance with PNIPAM, in IPN microgels the fingerprints of two different microscopic mechanisms are recognized. The first is similar to pure PNIPAM microgel: when the microgel particle goes towards the shrunken state, the number of water molecules surrounding the methyl group is reduced and the frequency of CH$_3$ stretching abruptly decreases.  
The latter is evidenced by an additional bump in the temperature dependence of the CH$_3$ frequency. This behaviour is peculiar of IPN microgel and is due to the presence of PAAc that enhances microgel hydrophobicity and limits the microgel shrinking.
\\ Raman spectroscopy is therefore a particularly suitable probe for investigating the microscopic underlying interactions in these complex 
systems, showing that hydrophobicity plays a crucial role not only in the swelling behaviour but also in the aggregation phenomena.

\subsection*{Bibliography}

\bibliographystyle{unsrt}\biboptions{sort&compress}

\end{document}